\documentclass[prl,twocolumn,showpacs]{revtex4}
\usepackage{graphicx}
\usepackage{times}

\begin{document} 
\title{Quantum homodyne tomography of a two-photon Fock state}

\author{Alexei Ourjoumtsev}
\author{Rosa Tualle-Brouri}
\author{Philippe Grangier}
\affiliation{Laboratoire Charles Fabry de l'Institut d'Optique, CNRS UMR 8501, 91403 Orsay, France}
\date{\today}

\begin{abstract}
 We present a continuous-variable experimental analysis of a two-photon Fock state of free-propagating
light. This state is obtained from a pulsed non-degenerate parametric amplifier, which produces two
intensity-correlated twin beams. Counting two photons in one beam projects the other beam in the desired
two-photon Fock state, which is analyzed by using a pulsed homodyne detection. The Wigner function of the
measured state is clearly negative. We developed a detailed analytic model which allows a fast and
efficient analysis of the experimental results.
\end{abstract}
\maketitle

Quantum properties of light beams can be described in terms of amplitude
and phase or, in Cartesian coordinates, in terms of the ``quadrature
components" of the quantized electric field, associated with
non-commuting operators $\hat{x}$ and $\hat{p}$.
The corresponding observables, often called ``quantum continuous variables", are analogous to the position and the momentum of a particle, and
from Heisenberg's inequalities they cannot be determined simultaneously
with an infinite precision. As a consequence, one cannot define a proper
phase-space distribution $\Pi(x,p)$ for the electric field, but rather a
quasidistribution $W(x,p)$ called the Wigner function. This function
can be reconstructed by quantum homodyne tomography \cite{VogelTomography}, which consists
in measuring several quadratures $\hat{x}_{\theta}=\hat{x} \; \cos\theta+\hat{p} \; \sin\theta$ with a homodyne
detection, and applying an inverse Radon transform. 

The most conspicuous
property of the Wigner function is that it may take negative values
for specific quantum states, as a signature of their non-classical
nature. This is the case for Fock
states, which contain a well-defined number of photons. 
Such states can be generated by using ``twin" beams, which are produced by
optical parametric amplification, and which contain perfectly
correlated numbers of photons. Counting $n$
photons in one mode projects the other mode in a $n$-photon Fock state,
which can then be analyzed using homodyne tomography. This was recently demonstrated for $n=1$ \cite{Lvovsky1photon,Zavatta1photon}.
However, up to now this method
could not be applied for higher photon numbers, since the probability to
generate simultaneously more than one photon pair was extremely low.

In this Letter we present a detailed analysis of a
free-propagating light pulse prepared in a two-photon Fock state $(n=2)$. The
measured Wigner function presents a complex structure and takes negative
values. In addition to standard methods, we will also present a novel analytic 
model of the experiment,  allowing an in-depth
physical interpretation of the experimental results.

\begin{figure}[b]
\includegraphics[width=7cm]{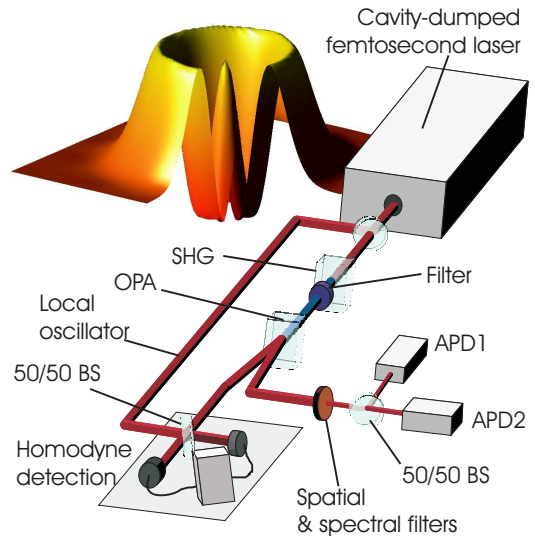}
\caption{Experimental setup, and Wigner function of the two-photon state propagating in the experiment
(corrected for homodyne detection losses, see text).}
\label{setup}
\end{figure}

\medskip
 Our experimental setup is presented on Fig.~\ref{setup}. A pulsed Ti-Sapphire laser
produces 180-femtosecond nearly Fourier-limited pulses with an energy of 40 nJ and a 800 kHz repetition rate
\cite{WengerDetHom}. The high pulse peak power allows us
to increase the pair production rate beyond what was available previously \cite{Lvovsky1photon,Zavatta1photon}.
The 850~ nm pulses are frequency-doubled [second harmonic generation (SHG)] by a single pass in a $100 \; \mu m$ thick
non-critically phase-matched potassium niobate (KNbO$_{3}$) crystal. The frequency-doubled beam pumps an
identical crystal used as an optical parametric amplifier (OPA), generating a two-mode squeezed state
\cite{PulsedEPR}. To align the setup, a probe beam is injected in the OPA with an angle of
$5^{\circ}$ to the pump direction. It allows to measure a classical phase-independent gain
$g=1.07$. The homodyne detection is aligned on the idler beam, whereas the signal beam, after spatial and
spectral filtering, is split between two avalanche photodiodes (APD) operating in a photon-counting regime. The
detection of a coincidence by the APDs means that at least two photon pairs were created in the OPA by the
same pulse. Since the gain $g$ is still relatively low the probability to create more than two pairs is small
in this case. Therefore, a coincidence detected by the APDs conditionally prepares a two-photon state in the
 idler beam. Single-photon states are conditioned by single APD events. The prepared states are analyzed by a homodyne detection
operating in a time-resolved regime. It samples each individual pulse, measuring one quadrature $X_{\theta}$ in phase with the
local oscillator. 

In previous $n=1$ state reconstruction experiments \cite{Lvovsky1photon,Zavatta1photon,Bertet1Photon}, 
it was generally admitted that the generated states are phase-independent. 
In our case, the production rate of single photons is very high, 
and we can record the full $n=1$ quadrature distribution in less than a second, during which phase
drifts are negligible. Therefore we did check experimentally that both the unconditional (thermal) and singly-conditional $(n=1)$
probability distributions do not depend on  $\theta$. Then it is quite reasonable to 
assume that this is also the case for the $n=2$ state, as it was done for the $n=1$ state in older experiments. 

\begin{figure}
\includegraphics[width=7cm, height=5cm]{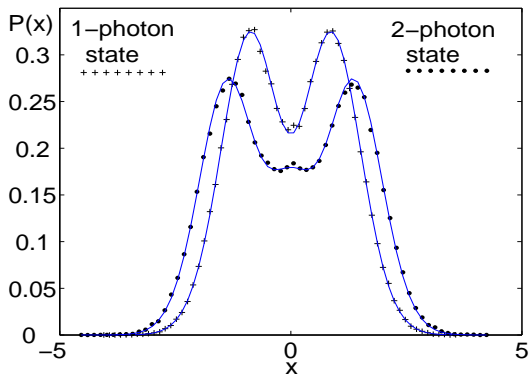}
\caption{Experimental quadrature measurements, and quadratures reconstructed using our model (see text)}
\label{quadratures}
\end{figure}

\begin{figure}
\includegraphics[width=4.2cm]{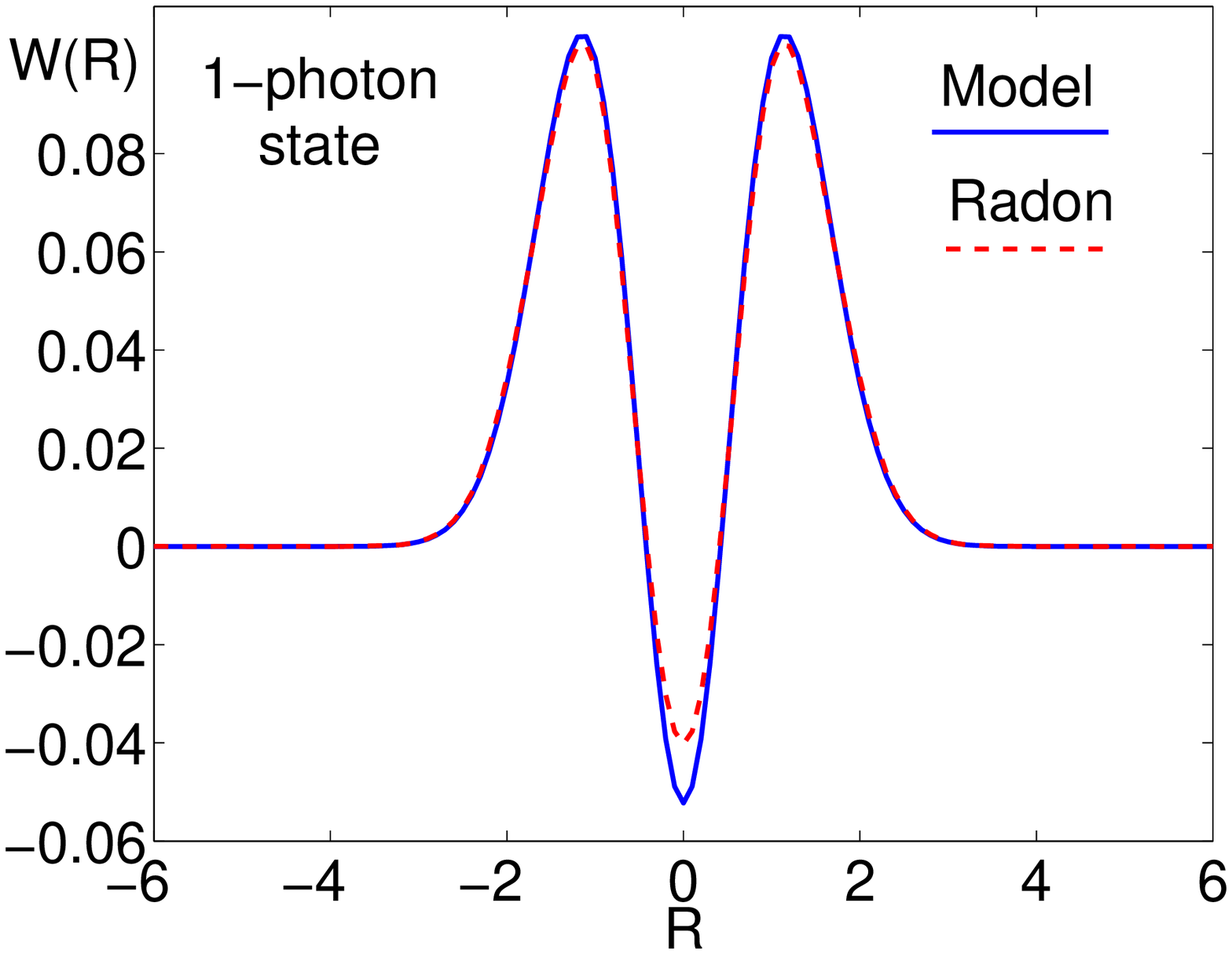}
\includegraphics[width=4.2cm]{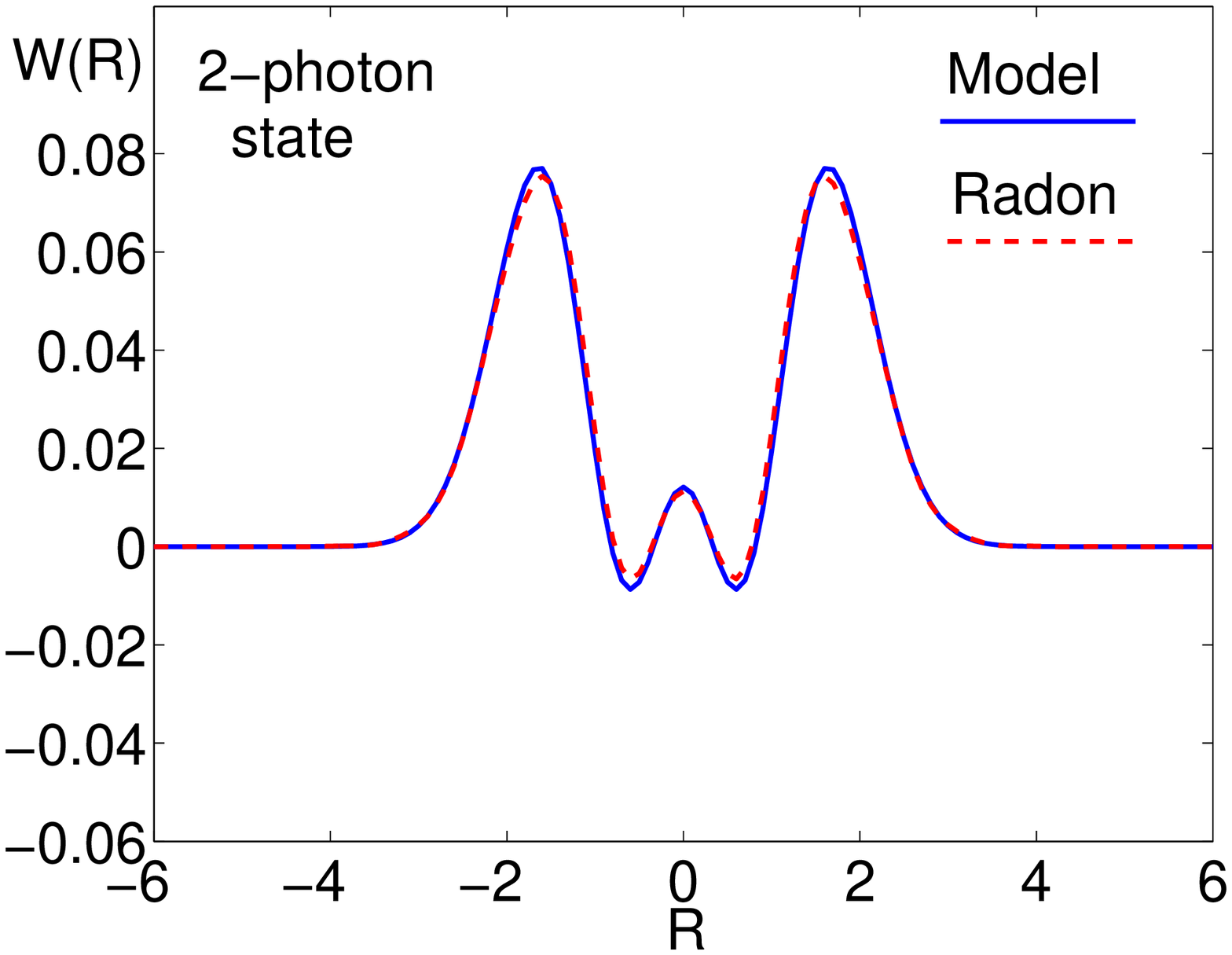}
\caption{Wigner functions of the measured uncorrected states, reconstructed by a
standard Radon transform, compared to those obtained using the model described in the text}
\label{Radon}
\end{figure}

\begin{figure}
\includegraphics[width=4cm]{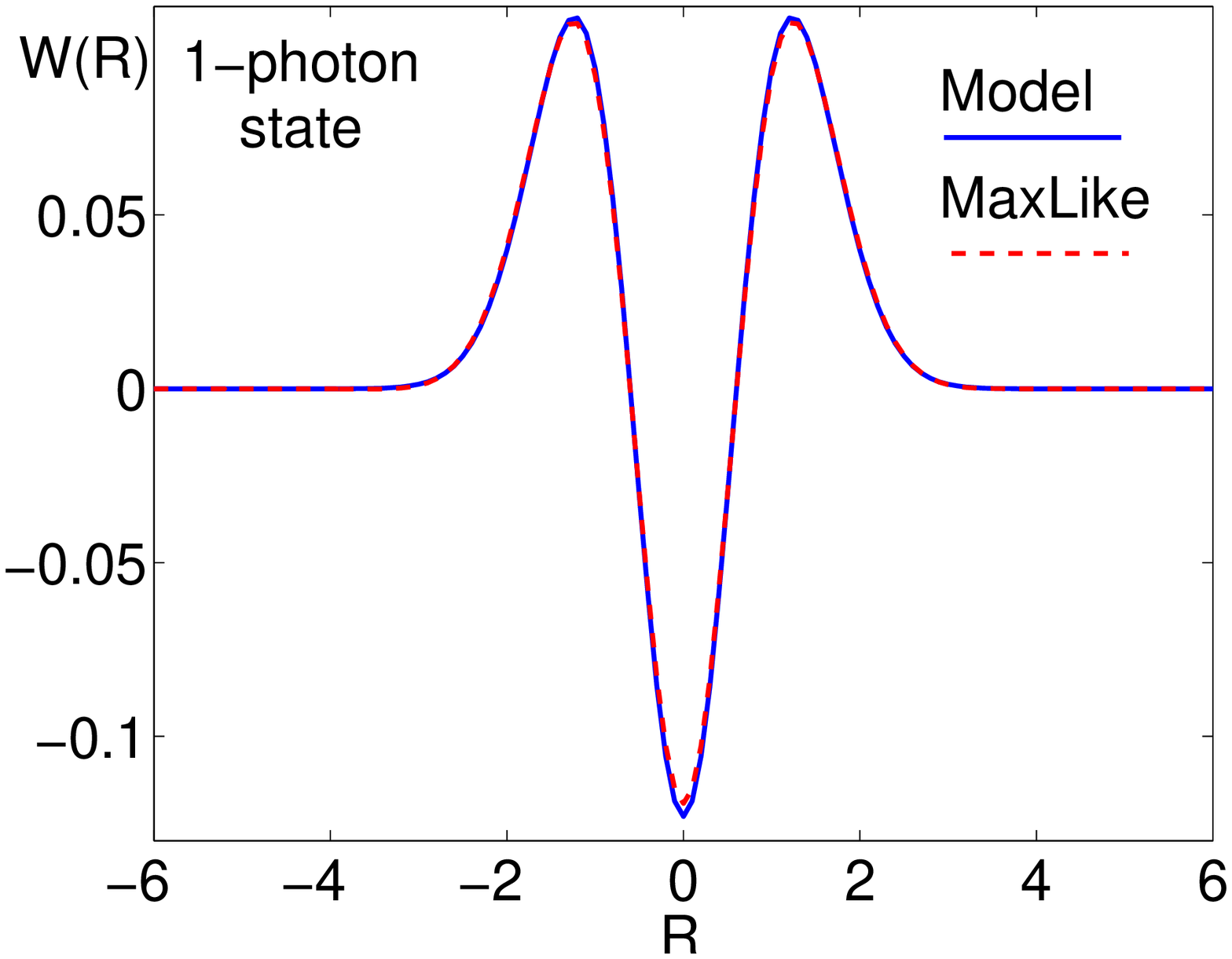}
\includegraphics[width=4cm]{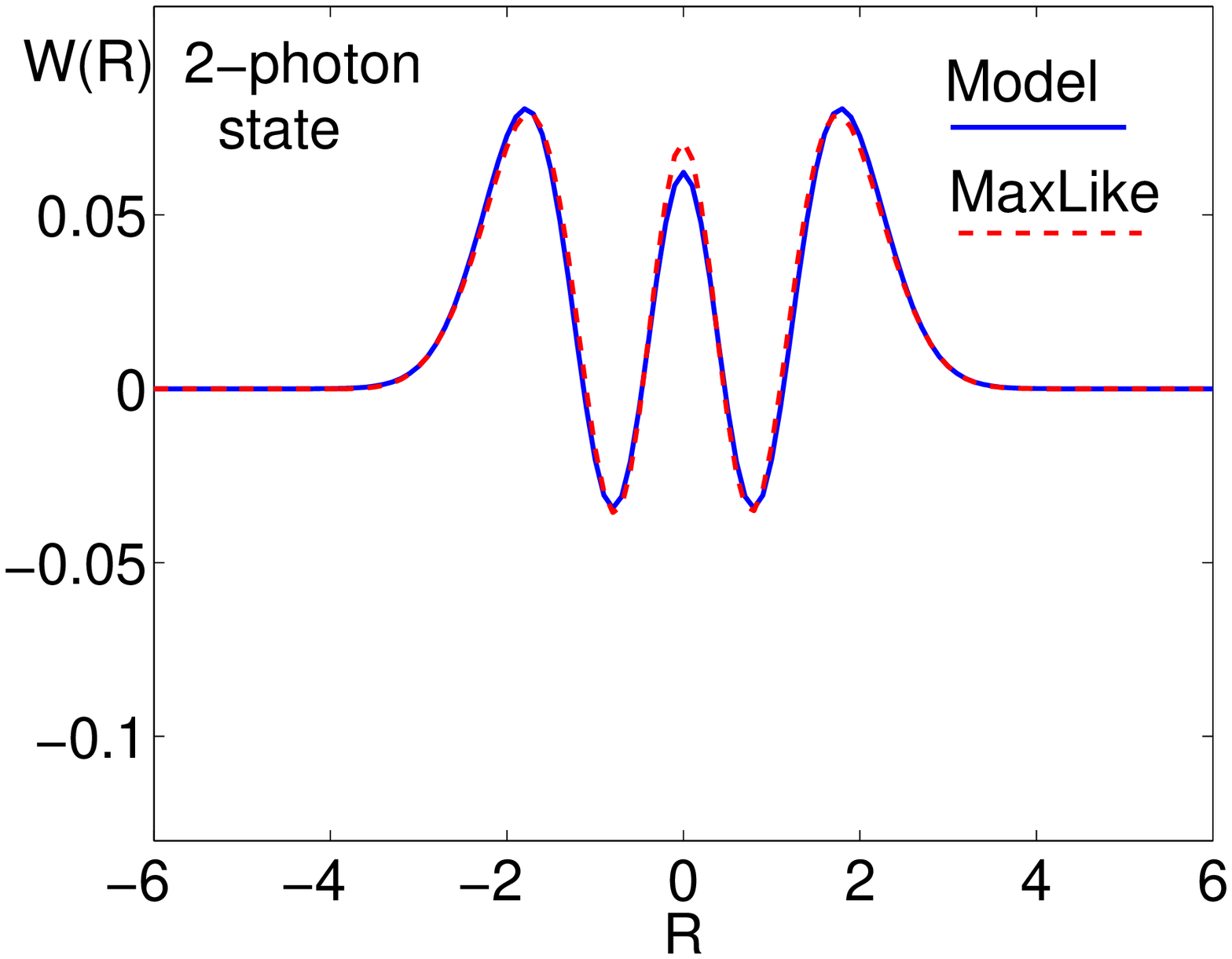}
\caption{Experimental Wigner functions corrected for losses in the homodyne detection, reconstructed by a
standard maximal-likelihood (MaxLike) method, compared to those obtained using the model described in the text}
\label{MaxLike}
\end{figure}

In a 2-hour experimental run we acquired $105.000$ homodyne data points conditioned on two-photon coincidences (40 seconds were
enough to acquire 180.000 single-photon events). Dividing the data into 64-bin histograms, we obtained the quadrature
distributions presented on Fig.~\ref{quadratures}. With a numerical Radon transform, we reconstructed the Wigner functions
associated with the measured states (see Fig.~\ref{Radon}), both clearly negative. Their minima and their values at the origin
are presented in Table~\ref{dataWig}. To determine the Wigner functions of the \textit{generated} states, presented on
Fig.~\ref{MaxLike}, we correct for the homodyne detection losses by using a standard maximal-likelihood (MaxLik) algorithm
\cite{FiurasekMaxLik,LvovskyMaxLik}, taking into account an independently measured homodyne efficiency $\eta=80\%$.

\begin{table}
  \caption{Critical values of the Wigner functions corresponding to the measured uncorrected data 
(Raw, obtained from the Radon transform),
to the state corrected for homodyne detection losses (Corrected, obtained from the MaxLik method) and to the ideal state (Ideal).}
\begin{ruledtabular}
\begin{tabular}{ccccc}
  & \multicolumn{2}{c}{\textbf{2 photons}} && \textbf{1 photon} \\
  &  $min(W_2)$            &  $W_2(0)$ &&  $min(W_1)=W_1(0)$\\
  \hline
Raw        & $-0.009 \pm 0.003$ & $0.012 \pm 0.003$&&$-0.052 \pm 0.003$\\
Corrected  & $-0.034 \pm 0.003$ & $0.062 \pm 0.003$&&$-0.123 \pm 0.003$\\
Ideal      & $-0.13$  & $0.32$ &&$-0.32$
\end{tabular}
\end{ruledtabular}
\label{dataWig}
\end{table}

\medskip
The negativity of the Wigner function can be rapidly lost with experimental imperfections. Above all, we must ensure that the prepared state belongs to the mode
analyzed by the homodyne detection. This modal overlap $\xi$ is decreased by the imperfections of
the filtering system, by the APD dark counts, and by the limited spectral and spatial qualities of the optical
beams. As a result, we may consider that the state is prepared in the right mode with a probability
$\xi$, and in an orthogonal mode with a probability $1-\xi$. A second source of decoherence is excess
noise in the OPA, producing uncorrelated photons. The actual OPA can be represented
by an ideal non-degenerate amplifier with a gain $g=\cosh^{2}(r)$, producing a pure two-mode squeezed state,
followed by two phase-independent amplifiers on signal and idler beams, each one with a gain
$h=\cosh^{2}(\gamma r)$, where $\gamma$ is the ratio between the undesired and the desired amplification
efficiencies (ideally $\gamma=0$).  Finally, the homodyne detection presents a finite efficiency $\eta$ and an
excess noise
$e$. From the measured optical transmission $\eta_t=97\%$, quantum detection efficiency $\eta_q=97.5\%$ and mode-matching
efficiency $\eta_m=92\%$, we estimate $\eta=\eta_t \eta_q \eta_m^2 = 80\%$. 
Since $\eta$ and $e$ are not involved in the preparation but only in the analysis of the state, we can correct for their effects in order to determine the actual Wigner function of the generated
state. The overall efficiency $\mu$ of the APD detection channel, although
rather low ($6\%$), is not a limitation in this experiment (see Appendix).

\medskip
In order to obtain a more physical analysis of our data, 
we have constructed a complete - but nevertheless simple - 
analytic model of the experiment (see Appendix). Apart from
predicting the performance of the setup, it allows to extract much more information from the experimental data than the
numerical methods presented above, although it is, of course, less general. It uses a generic parameterized 
expression of the Wigner function, derived in the Appendix, which accounts for all the experimental defects~: 
\begin{eqnarray}
W_{2}(x,p) &=& \frac{e^{-{\textstyle \frac{R^{2}}{\sigma^{2}}}}}{\pi \sigma^{2}} \left[(1-\delta)^{2}+2(1-\delta) \frac{\delta R^{2}}{\sigma^{2}} + \frac{\delta^{2} R^{4}}{2 \sigma^{4}} \right] \; \; \; \label{eqW2}\\
\nonumber \text{where}&& R^{2}= x^{2}+p^{2} \\
&&\sigma^{2} = 2\eta(h g-1)+1+e \label{eqSigma}\\
&&\delta = 2\xi\eta h^{2}g(g-1)/[\sigma^{2}(h g-1)] \label{eqDelta}
\end{eqnarray}

\noindent The associated quadrature distribution is described by
\begin{eqnarray}
P_{2}(x)  =  \frac{e^{-x^{2}/\sigma^{2}}}{\sqrt{\pi \sigma^{2}}}  \left[1-\delta+\frac{3\delta^{2}}{8} + \frac{4-3\delta}{2}\frac{\delta
x^{2}}{\sigma^{2}} +\frac{\delta^{2}x^{4}}{2\sigma^{4}}\right]
\end{eqnarray}
For the one-photon case, the same method leads to
\begin{eqnarray}
  W_{1}(x,p)& = & \frac{e^{-R^{2}/\sigma^{2}}}{\pi \sigma^{2}} \left[1-\delta+ \frac{\delta R^{2}}{\sigma^{2}}\right]
\label{eqW1}\\
 P_{1}(x)& = & \frac{e^{-x^{2}/\sigma^{2}}}{\sqrt{\pi \sigma^{2}}} \left[1-\frac{\delta}{2}+ \frac{\delta x^{2}}{\sigma^{2}}\right]
\end{eqnarray}
The density matrices of these states are diagonal in the Fock basis, the non-zero coefficients given by :
\begin{eqnarray}
\langle n \vert \rho_{2} \vert n \rangle  &=&  \frac{2 (\sigma^{2}-1)^{n-2}}{(\sigma^{2}+1)^{n+3}}
\left[S_{n}^{2}- 2n(n+1)\delta^{2}\sigma^{4} \right]\\
\langle n \vert \rho_{1} \vert n \rangle  &=&  2 S_{n} (\sigma^{2}-1)^{n-1}/(\sigma^{2}+1)^{n+2}\\
\langle n \vert \rho_{0} \vert n \rangle  &=&  2 (\sigma^{2}-1)^{n}/(\sigma^{2}+1)^{n+1}
\end{eqnarray}

where $S_{n}  = \sigma^{4}(1-\delta)+\sigma^{2}\delta(1+2n)-1$, 
and $\rho_{0}$ corresponds to the thermal unconditioned state (obtained by taking $\delta = 0$ in 
any of the above equations).

These states are completely described by the two same parameters $\sigma^{2}$ and $\delta$.
 Here $\sigma^{2}$ is simply the variance of the non-conditioned gaussian thermal state. The non-classicality of the conditioned states is
determined by $\delta$, which varies between $0$ for a non-conditioned state and $2$ for the ideal case. When
$\delta>1$, both $W_{1}$ and $W_{2}$ become negative, and a central peak appears on $W_{2}$. These parameters, very useful to optimize the experiment, can be directly extracted from the second and fourth moments of the measured distributions :

\medskip
\begin{tabular*}{0.4\textwidth}{@{\extracolsep{\fill}} rcllrcl}
\multicolumn{3}{c}{1 photon} && \multicolumn{3}{c}{2 photons} \\
\hline
 $\langle x^{2} \rangle_{1}$ & $=$ & $\sigma^{2}(1+\delta)/2\;$  &&  $\; \langle x^{2} \rangle_{2}$ & $=$ &
$\sigma^{2}(1+2\delta)/2$\\
 $\langle x^{4} \rangle_{1}$ & $=$& $ 3\sigma^{4}(1+2\delta)/4\;$  &&  $\; \langle x^{4} \rangle_{2}$ & $=$ &
$3\sigma^{4}(1+4\delta+\delta^{2})/4$
\end{tabular*}
\medskip

We used one-photon conditioning during the optimization, so that $\sigma^{2}$ and $\delta$
could be determined in a few seconds, $300$ times faster than in the two-photon case. The two-photon state, described in
 principle by the same parameters, was ``automatically" optimized in this process. We found that the values deduced from single
and two-photon state tomographies are exactly the same for 
$\sigma^{2}$, and differ by less than two percent for $\delta$.

In addition, the quadratures reconstructed using the parameters $\sigma^{2}$ and $\delta$ extracted from raw data are in
excellent agreement with the measurements (see Fig.~\ref{quadratures}), and the reconstructed Wigner functions of the measured
states are very close to those obtained by the Radon transform (Fig.~\ref{Radon}). 
Equations \ref{eqSigma} and \ref{eqDelta} also allow to determine the modal overlap $\xi$ and the excess gain parameter $\gamma$. 
The obtained values ($\xi=0.9$ and $\gamma = 0.4$) are fully compatible with experimental evaluations, which are difficult to do
but were carried out by using independent classical amplification and photon counting techniques.

\begin{figure}
\includegraphics[width=2.8cm]{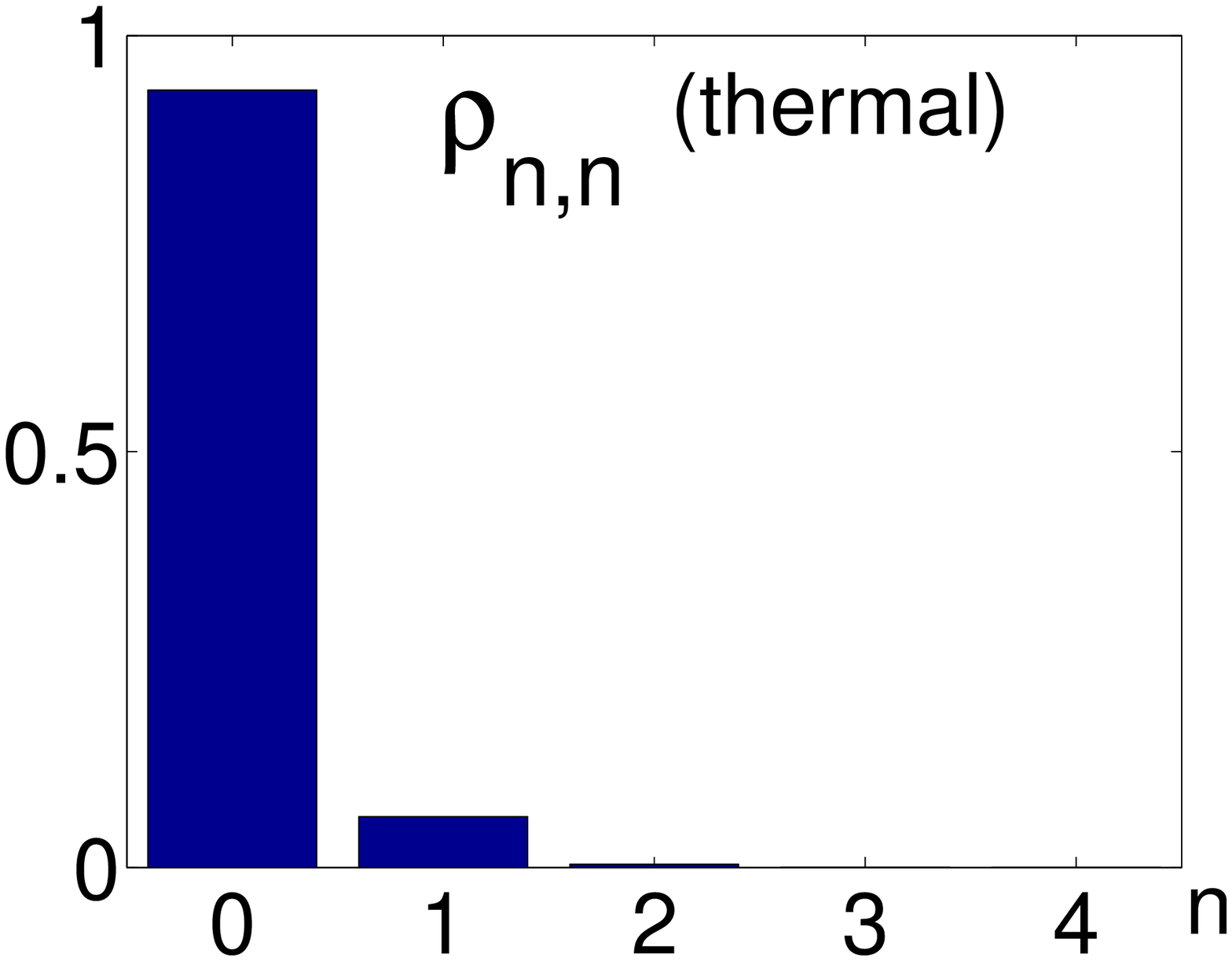}
\includegraphics[width=2.8cm]{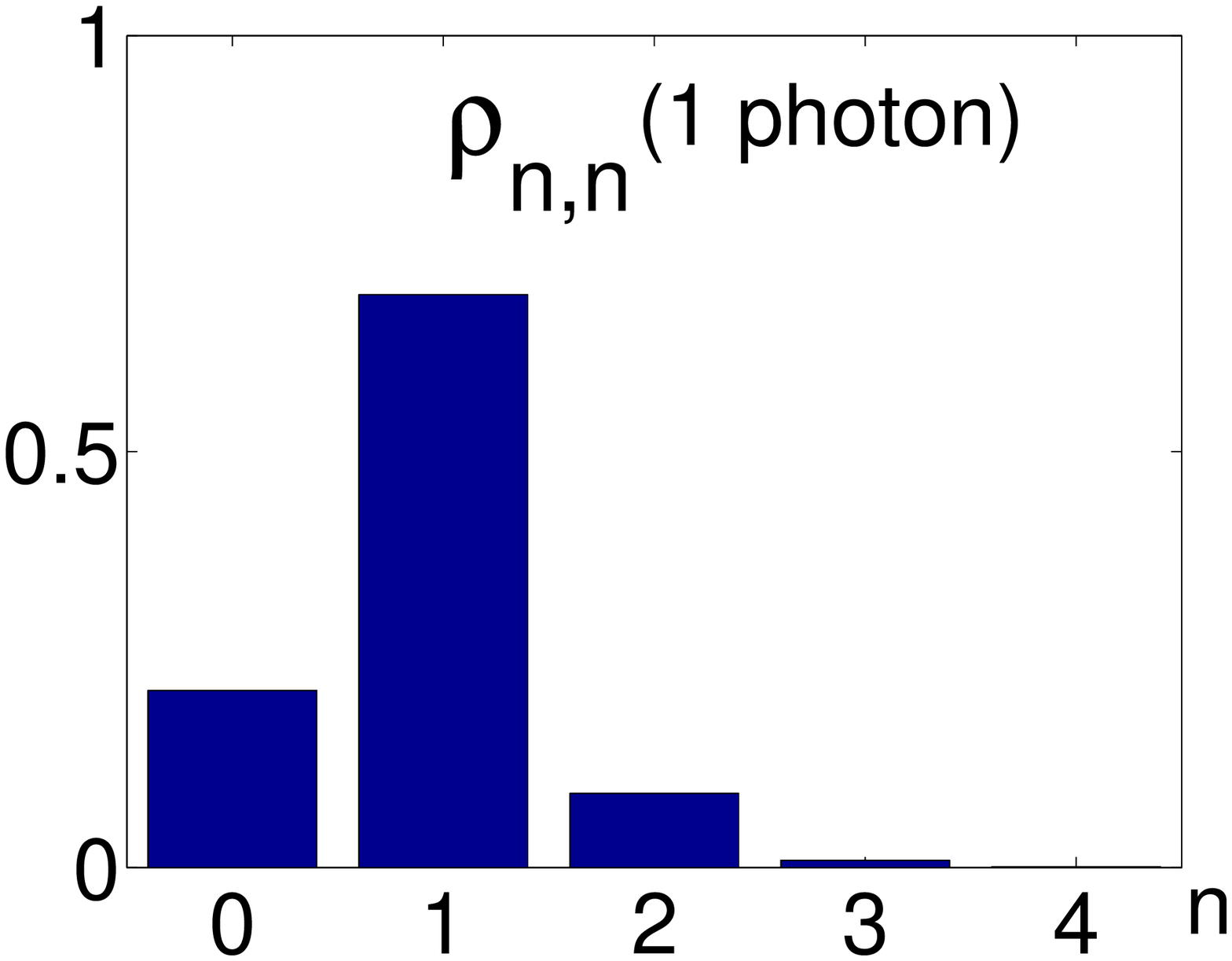}
\includegraphics[width=2.8cm]{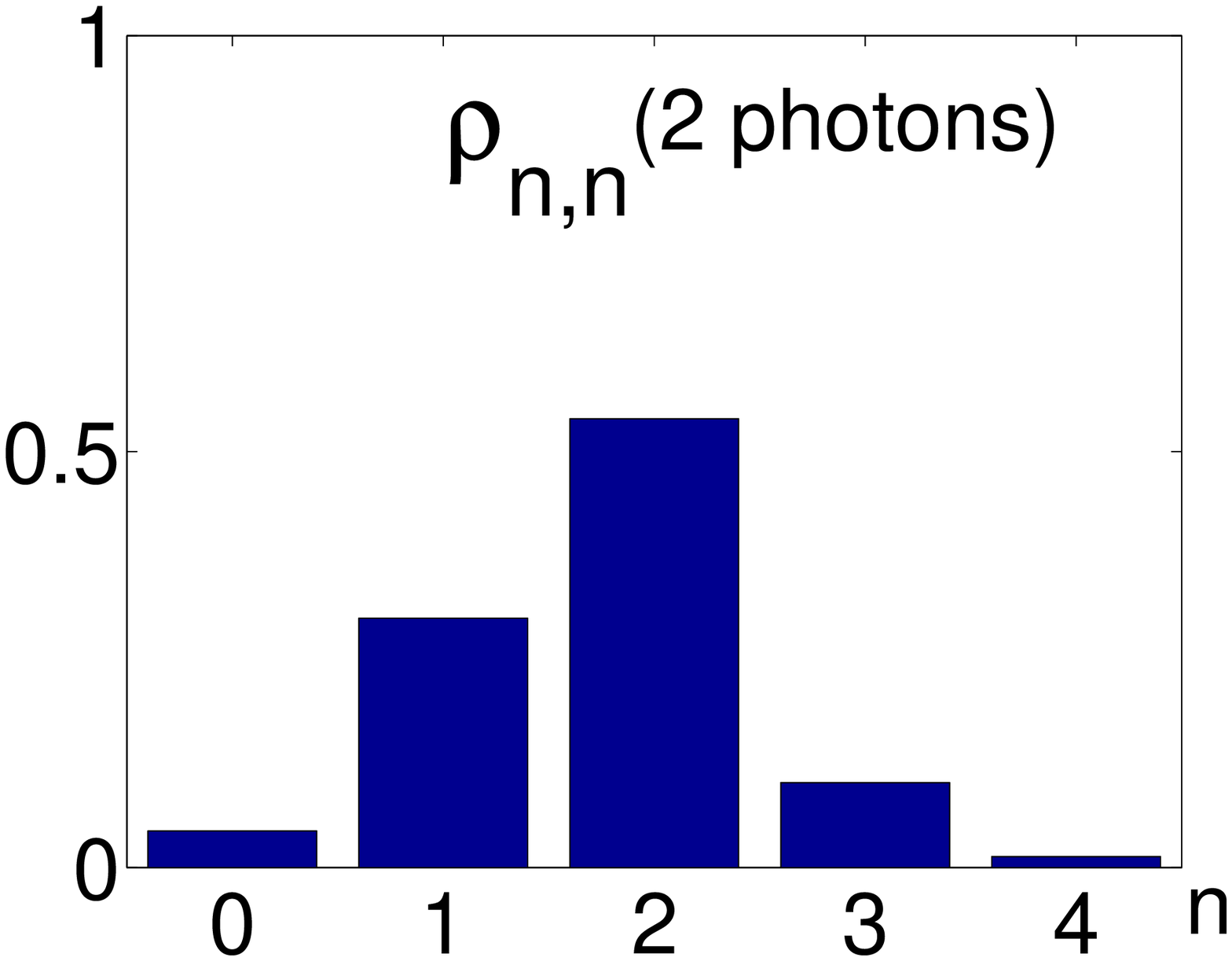}
\caption{Main density matrix coefficients of the states conditioned on $0$, $1$ and $2$ photons (corrected for detection losses).}
\label{matrix}
\end{figure}

 Since the results obtained with this method appear to be completely consistent, both within themselves and with independant measurements, we can assume
that the Wigner function of the
\textit{generated} state, which we would measure with an ideal homodyne detection, can be simply calculated by taking $\eta=1$ and $e=0$ in our
expressions, keeping all other parameters unchanged. The obtained results are again in good agreement with those provided by the maximal-likelihood method,
as shown on Fig.~\ref{MaxLike}. The main density matrix coefficients of the generated states are represented on Fig.~\ref{matrix}. This gives confidence
that our method provides a very fast and reliable way to interpret the experimental data, which is more ``constrained" than the Radon transform, but
also much closer to the physics of the experiment. 

 The present experimental and theoretical results demonstrate simple techniques to generate and analyze sophisticated non-classical states of
propagating light fields, which have been considered almost out of experimental reach during many years. Similar methods can be used to
create photon-subtracted entangled states with two-mode negative Wigner functions, which should improve the fidelity in teleportation experiments
\cite{OpatrnyCondTeleport,CochraneCondTeleport,OlivaresCondTeleport}, and allow to implement loophole-free Bell tests
\cite{BellCerf,BellCar}. The avenue of manipulating negative Wigner functions now seems clearly open for quantum communications. 

\section*{APPENDIX}

\begin{figure}
\includegraphics[width=8cm,height=2.7cm]{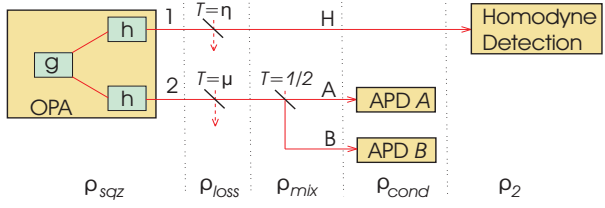}
\caption{Modelling of the experiment.}
\label{model}
\end{figure}

The model for the experiment is represented on Fig. \ref{model}.
The OPA produces a two-mode noisy squeezed state with a density matrix $\rho_{sqz}$ associated with a Wigner
function
\begin{eqnarray}
\lefteqn{W_{sqz}(x_{1},p_{1},x_{2},p_{2}) =} \\
 \nonumber  & &
= \frac{\exp\left(-\frac{(x_{1}-x_{2})^{2}+(p_{1}+p_{2})^{2}}{(hs+h-1)}-\frac{(x_{1}+x_{2})^{2}+(p_{1}-p_{2})^{2}}{(h/s+h-1)}\right)}{\pi^{2}
(hs+h-1)(h/s+h-1)}
\end{eqnarray}
where $s=e^{-2 r}$ is the two-mode variance squeezing factor associated with a gain $g=\cosh^{2}(r)$, and
$h=\cosh^{2}(\gamma r)$ is the excess gain. The mode $1$ is directed towards the homodyne detection, whereas the mode $2$ is
sent into the conditioning channel. The homodyne losses can be represented by mixing the mode $1$ with vacuum on a
beam splitter (BS) with a transmission  $T=\eta$. Since we are only interested in the transmitted mode $H$, we trace over
the reflected mode to obtain the resulting density matrix. The same holds for the APD losses, with a transmission
$T=\mu$. 

The resulting Wigner function $W_{loss}$ is calculated by convolution of $W_{sqz}$ with the Wigner functions
$W_{vac}$ of two vacuum modes, using $W_{vac}(x,p) = \exp(-x^{2}-p^{2})/\pi$. Then, the mode transmitted through the APD
channel is mixed with another vacuum mode on a 50/50 beamsplitter, producing a density matrix $\rho_{mix}$ involving three
modes $H$, $A$ and $B$, and associated with the Wigner function~: 
\begin{center}
\begin{tabular}{ccc}
$W_{mix}$ &=& $W_{loss}\left(x_{H},p_{H},\frac{x_{A}+x_{B}}{\sqrt{2}},\frac{p_{A}+p_{B}}{\sqrt{2}}\right)$\\
& & $\times W_{vac}\left(\frac{x_{A}-x_{B}}{\sqrt{2}},\frac{p_{A}-p_{B}}{\sqrt{2}}\right)$
\end{tabular}
\end{center}
\noindent

\noindent The modes $A$ and $B$ are detected by the APDs $A$ and $B$, which realize respectively the projective measurements  
$\Pi_{A,B}=Id-\vert 0_{A,B}\rangle \langle 0_{A,B} \vert$ with a probability $\xi$ (``matched clicks"),
and $\Pi_{0}=Id$ with a probability $1-\xi$ (``unmatched clicks"). The density matrix becomes
\begin{eqnarray}
\nonumber \rho_{cond}&=& N_{2} \xi^{2} \Pi_{A} \Pi_{B} \rho_{mix} \Pi_{A} \Pi_{B} + (1-\xi)^{2} \rho_{mix}\\
\nonumber && + N_{1} \xi (1-\xi ) (\Pi_{A} \rho_{mix} \Pi_{A} + \Pi_{B} \rho_{mix} \Pi_{B})
\end{eqnarray}
where $N_{1}=1/Tr(\Pi_{A} \rho_{mix})=1/Tr(\Pi_{B} \rho_{mix})$ and $N_{2}=1/Tr(\Pi_{A} \Pi_{B} \rho_{mix})$.
Finally, the density matrix of the measured two-photon state is obtained by tracing out the two APD modes $A$ and $B$~:
\begin{eqnarray}
\nonumber \rho_{2}&=& Tr_{A,B} \; \rho_{cond}\\
\nonumber &=&  \left[N_{2} \xi^{2} + 2 N_{1} \xi (1-\xi ) +(1-\xi )^{2} \right] Tr_{A,B} \; \rho_{mix} \\
\nonumber && -\left[N_{2} \xi^{2} + N_{1} \xi (1-\xi ) \right] Tr_{B} \langle 0_{A} \vert \rho_{mix} \vert 0_{A} \rangle \\
\nonumber && -\left[N_{2} \xi^{2} + N_{1} \xi (1-\xi ) \right] Tr_{A}\langle 0_{B} \vert \rho_{mix} \vert 0_{B} \rangle \\
 && +N_{2} \xi^{2} \langle 0_{A} 0_{B} \vert \rho_{mix}\vert 0_{A} 0_{B} \rangle
\end{eqnarray}
The associated Wigner function can be calculated using
\begin{center}
\begin{tabular}{ccc}
$Tr_{K}W_{mix}$  & = &  $\int{W_{mix}}dx_{K}dp_{K}$\\
$\langle 0_{K} \vert W_{mix} \vert 0_{K} \rangle$  & = &  $2\pi\int{W_{mix}W_{vac}}dx_{K}dp_{K}$
\end{tabular}
\end{center}
\noindent
where $K = A, B$. As expected, it has no definite phase and depends only on $R^{2}=x_{H}^{2}+p_{H}^{2}$. It has the form
\begin{eqnarray}
W_{2}& = & \frac{\alpha e^{-{\textstyle \frac{R^{2}}{\sigma_{2}^{2}}}} }{\pi \sigma_{2}^{2}}  - \frac{\beta
e^{-{\textstyle \frac{R^{2}}{\sigma_{1}^{2}}}} }{\pi \sigma_{1}^{2}}  + \frac{(1-\alpha+\beta) e^{-{\textstyle \frac{R^{2}}{\sigma^{2}}}} }{\pi
\sigma^{2}} \label{eqW2exact}
\end{eqnarray}

\noindent
where $\alpha$, $\beta$ and $\sigma_{i}$ are functions of the parameters above.
This linear combination of gaussian functions looks quite simple, 
but $\alpha$ and $\beta$ diverge
when the OPA gain or the APD efficiency are small, which is our case. This leads to
numerical instabilities when this expression is used for data analysis. To avoid this problem one can simply take the
limit $\mu \rightarrow 0$ in eq.~(\ref{eqW2exact}), obtaining eq.~(\ref{eqW2}) quoted in the main text above. 
In our range of parameters, these two equations are numerically indistinguishable.

\begin{acknowledgments}
This work is supported by EU program COVAQIAL.
\end{acknowledgments}

\bibliography{Ourjoumt_2PhotonTomographyV2}

\end{document}